\documentclass[10pt,journal,compsoc]{IEEEtran}

\usepackage{ifpdf}

\ifCLASSOPTIONcompsoc
\usepackage[nocompress]{cite}
\else
\usepackage{cite}
\fi

\ifCLASSINFOpdf
\usepackage[pdftex]{graphicx}
\else
\fi

\usepackage{url}

\begin{document}

%
\title{Towards A Methodology and Framework for Workflow-Driven Team Science}
%
%
%
%

\author{Ilkay~Altintas,~\IEEEmembership{Member,~IEEE,}
        Shweta~Purawat,
        Daniel~Crawl,
        Alok~Singh,
        and~Kyle Marcus
\IEEEcompsocitemizethanks{\IEEEcompsocthanksitem I. Altintas, S. Purawat, D. Crawl, A. Singh and K. Marcus were with the San Diego Supercomputer Center at the University of California, San Diego while conducting the work towards this manuscript.\protect\\
E-mail: ialtintas@ucsd.edu}
}


\IEEEtitleabstractindextext{%
\begin{abstract}
Scientific workflows are powerful tools for management of scalable experiments, often composed of complex tasks running on distributed resources. Existing cyberinfrastructure provides components that can be utilized within repeatable workflows. However, data and computing advances continuously change the way scientific workflows get developed and executed, pushing the scientific activity to be more data-driven, heterogeneous and collaborative. Workflow development today depends on the effective collaboration and communication of a cross-disciplinary team, not only with humans but also with analytical systems and infrastructure. This paper presents a collaboration-centered reference architecture to extend workflow systems with dynamic, predictable and programmable interfaces to systems and infrastructure while bridging the exploratory and scalable activities in the scientific process. We also present a conceptual design towards the development of methodologies and tools for effective workflow-driven collaborations, namely the \textit{PPoDS} methodology and  the \textit{SmartFlows} Toolkit for smart utilization of workflows in a rapidly evolving cyberinfrastructure ecosystem.
\end{abstract}

}

\maketitle

\IEEEdisplaynontitleabstractindextext

%
\IEEEpeerreviewmaketitle

\IEEEraisesectionheading{\section{Introduction}\label{sec:introduction}}

\IEEEPARstart{O}{ver} the last two decades, scientific workflow systems, e.g., \cite{Kepler, Kepler2, Pegasus, Taverna, Swift}, have matured as powerful tools for computational data scientists to perform scalable experiments, often composed of complex tasks for data management and executable algorithms. They have especially been useful for resource allocation, task scheduling, performance optimization, and static coordination of tasks on a potentially heterogeneous set of resources \cite{sciwf}. Additionally, most scientific workflow systems today provide capabilities for provenance tracking, repeatability and partial reproducibility support \cite{provswf}. As a complement to the workflow capabilities, existing cyberinfrastructure provides powerful components that can be utilized as building blocks within workflows to translate the newest advances into impactful, repeatable solutions that can execute at scale. However, the last decade has also brought unprecedented data and computing advances that changed the way scientific workflows get developed and executed, pushing the scientific activity to be even more data-driven, heterogeneous and collaborative \cite{futurewf}.

Today’s computing has diverse workload characteristics spanning high-performance computing, high-throughput computing and big data analytics. The traditional supercomputing applications are stronger than ever on their way to embrace exascale computing capacity. As our ability to collect data in real-time from internet-of-things has improved, the demand to process such data at scale has increased and requires big data processing capabilities. We observe a growing number of applications, including smart cities, precision medicine, energy management and smart manufacturing, that require a combination of advanced data analytics with traditional modeling and simulations. In addition, thanks to the advances in new computer architectures, most scientific codes are ported for special environments, e.g., GPUs. There is also an increasing demand for computing from scientific disciplines like social sciences which were not traditionally seen as supercomputing disciplines. In fact, every domain of science and engineering today can take advantage of big data and computing. A challenge for today’s computing architectures is the ability to respond to such heterogeneous needs and lowering the barriers to computing for long tail researchers as well as supporting the most cutting-edge computing applications.

On the software side, we observe many new ways to manage big data and high-performance storage as well as new forms of data integrity technologies, e.g., blockchain. Use of analytical and big data frameworks, e.g., Spark \cite{spark} and Keras (keras.io), are common in individual machine learning applications and as a part of integrated data-driven scientific simulations. Such heterogeneous capability in computing and software brings with it the need for software systems that can coordinate applications across different scales of computing, data and networking needs. A number of software innovations like cluster virtualization and container technologies, e.g., Docker \cite{docker} and Singularity \cite{singularity}, increased the portability of these software frameworks and environments, making it possible to turn any executable to run as a service on multiple platforms. Kubernetes \cite{kubernetes} has emerged as a dynamic container and resource management platform that can automate the configuration and orchestration of computing resources for varying workloads. Gateways \cite{gateways}, Jupyter notebooks \cite{jupyter} and similar enabling web and mobile interface have lowered the barriers for many more to access data on the fly and take advantage of computing. 

All these make workflows even more needed at the converged application level to enable communications with data and computing middleware, while optimizing resources and dynamically adapting to the changes during the execution of integrated applications. Workflows provide an ideal programming model for deployment of computational and data science applications on all scales of computing and provide a platform for system integration of data, modeling tools and computing while making the applications reusable and reproducible. They make it possible to manage dynamic-data driven applications and decision support using advances in big data platforms and on-demand computing systems, e.g., dynamic data-driven fire behavior modeling in real-time \cite{firemap}. Moreover, there is a new opportunity here for workflows to become even more useful and more aligned with the way teams of scientists collaborate and develop integrated applications.

\textbf{Contributions. }The discussion in this paper lies in the heart of the above-mentioned needs and opportunities for workflows. Starting with the question ``can there be a methodology to make workflows a systematic part of the collaborative scientific process?'' and tackling the problem of ``what would a toolkit look like for optimizing workflow effectivity from multiple perspectives within a team?'', we present a new methodology and set of tools for team science and intelligent end-to-end workflow development. Specific contributions we present are:

\begin{enumerate}
\item an introduction to the conceptual PPoDS methodology for collaborative metric-based workflow design,
\item a framework design for measuring and testing exploratory workflows using the PPoDS metrics, 
\item a design for capturing data during exploratory workflow development to make intelligent scalability and steering possible,
\item an introduction to the SmartFlows Toolkit for real-time data collection, benchmarks and intelligence for smart workflow execution, and
\item a collaboration-centered reference architecture using contributions 1-4 to extend workflow systems with dynamic, predictable and programmable interfaces to teams, systems and scalable infrastructure while bridging the exploratory and scalable activities in the scientific process.
\end{enumerate}

\textbf{Outline. }The rest of this paper is organized as follows. In Section 2, we introduce the new team science and a reference architecture to make it effective (contribution 5). Section 3 introduces PPoDS Methodology and our current prototype for PPoDS explore-to-scale tools within the NSF CHASE-CI (contributions 1, 2 and 3). In Section 4, we introduce the SmartFlows Toolkit (Contribution 4). We review background work in Section 5 and conclude in Section 6.

\section{The New Computational Team Science}

\begin{figure}
\centering
\includegraphics[width=0.48\textwidth]{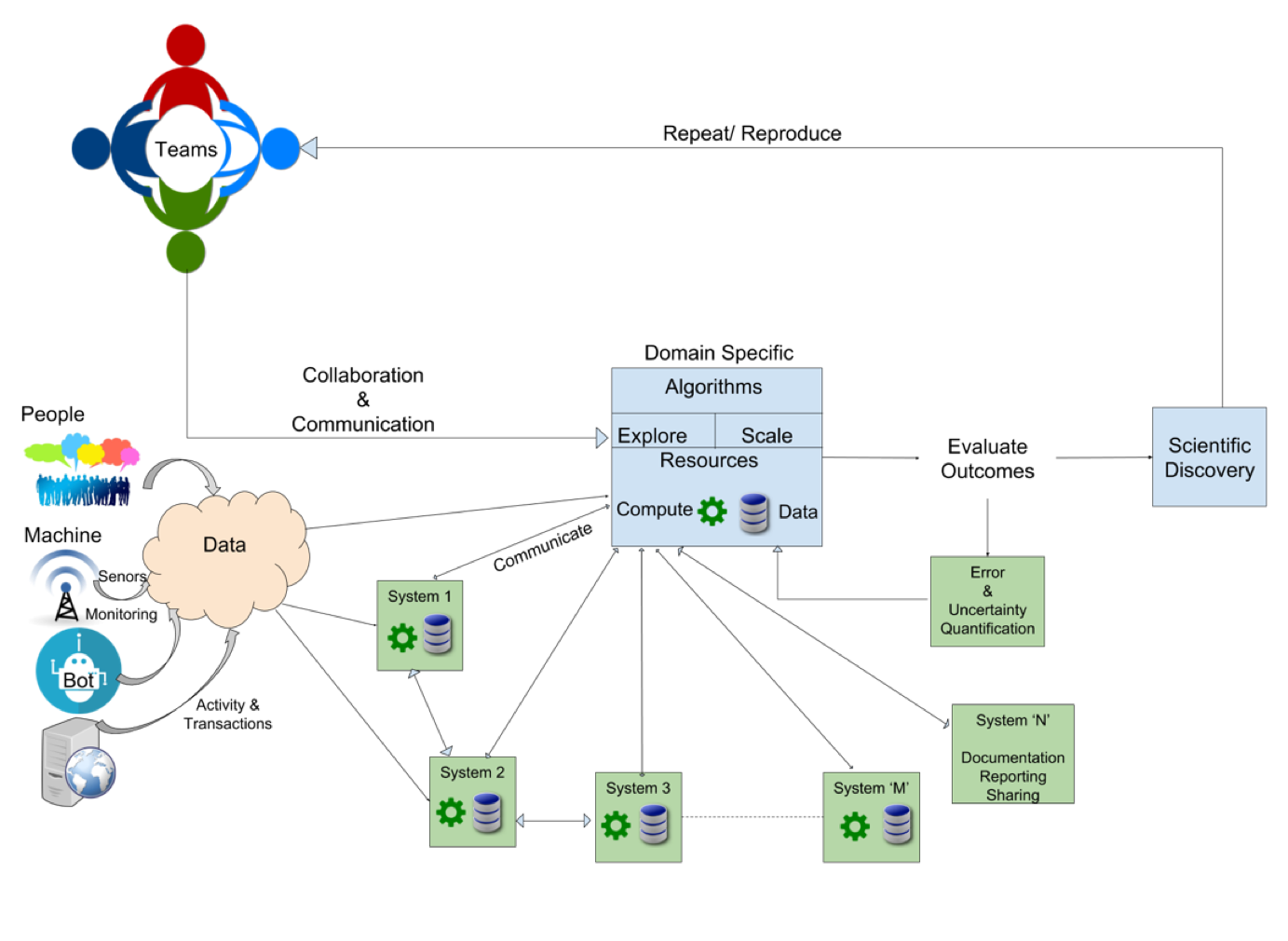}
\vspace{-12pt}
\caption{Collaborative team science from exploration to scalability to discovery.}
\label{fig:teamscience}
\vspace{-16pt}
\end{figure}   

As the complexity of scientific research grow to tackle the biggest problems of our time, the complexity of the tasks that need to be accomplished for a discovery also grow. Solutions to grand challenges of today require collaborative efforts of cross-disciplinary teams. The National Research Council report on the science of team science \cite{NASSOS} defines scientific collaboration as ``...research conducted by
more than one individual in an interdependent fashion, including research conducted by small teams and larger groups.'' In a computational and data-driven world,  in addition to conducting scientific analysis, each individual in a collaborative team does some of many other tasks. These tasks include execution of self-developed or community developed scientific analysis, modeling and simulation tools, interaction with many scales of computing, integration of big and small experimental historical or real-time datasets, development of methods to manage and interpret data, implementing communication and visualization dashboards, and managing data during and after its active period within the collaborative study. Such variety of individual efforts requires the effective collaboration and communication of a multi-disciplinary data science team with complementary scientific and technological expertise, not only with humans but also with analytical systems and infrastructure. 

Figure~\ref{fig:teamscience} illustrates a typical collaborative scientific research activity. A team of scientists with cross-disciplinary expertise collaborate and communicate to solve a problem. They explore  historical and real-time data, using storage, networking and computing resources available to them. Although the exploratory tasks may require forms of scale, these tasks are generally less demanding for resources, but very helpful in development of the approach and algorithms within the solution. Once the research methods are agreed upon, there is often a need for a more scalable execution of the solution that can lead to discoveries after careful evaluation of the research methods and outcomes. This scalable process often involves multiple steps with a need for coordination and requires repeatability. Over the last decade, we have witnessed many examples of workflow utilization for scalable process coordination and reproducibility in a wide range of scientific collaborations as an integral part of collaborative community cyberinfrastructure from physics \cite{PegasusLIGO} to wildfires \cite{KeplerWIFIRE} to chemistry \cite{mdwf}.

Another big challenge in the collaborative scientific process is keeping the link between exploratory activities and scalable process management. Often, after the exploratory activities, a different part of the team reengineers the developed methods for scale, making the iteration slower and reproducibility difficult. We argue that what we learn about the infrastructure resources, data management needs and algorithms in the exploratory analysis is key to the scalability process. Automating data collection in a way we can analyze and use as insight towards the scalability of the same process is currently rarely done or done in an ad-hoc fashion. Dynamic, predictable and programmable interfaces to exploratory systems and scalable infrastructure is key to building effective systems that can bridge the exploratory and scalable activities in the scientific process. 

\subsection{An Intelligent Workflow Framework for Team Science}

\begin{figure}
\centering
\includegraphics[width=0.48\textwidth]{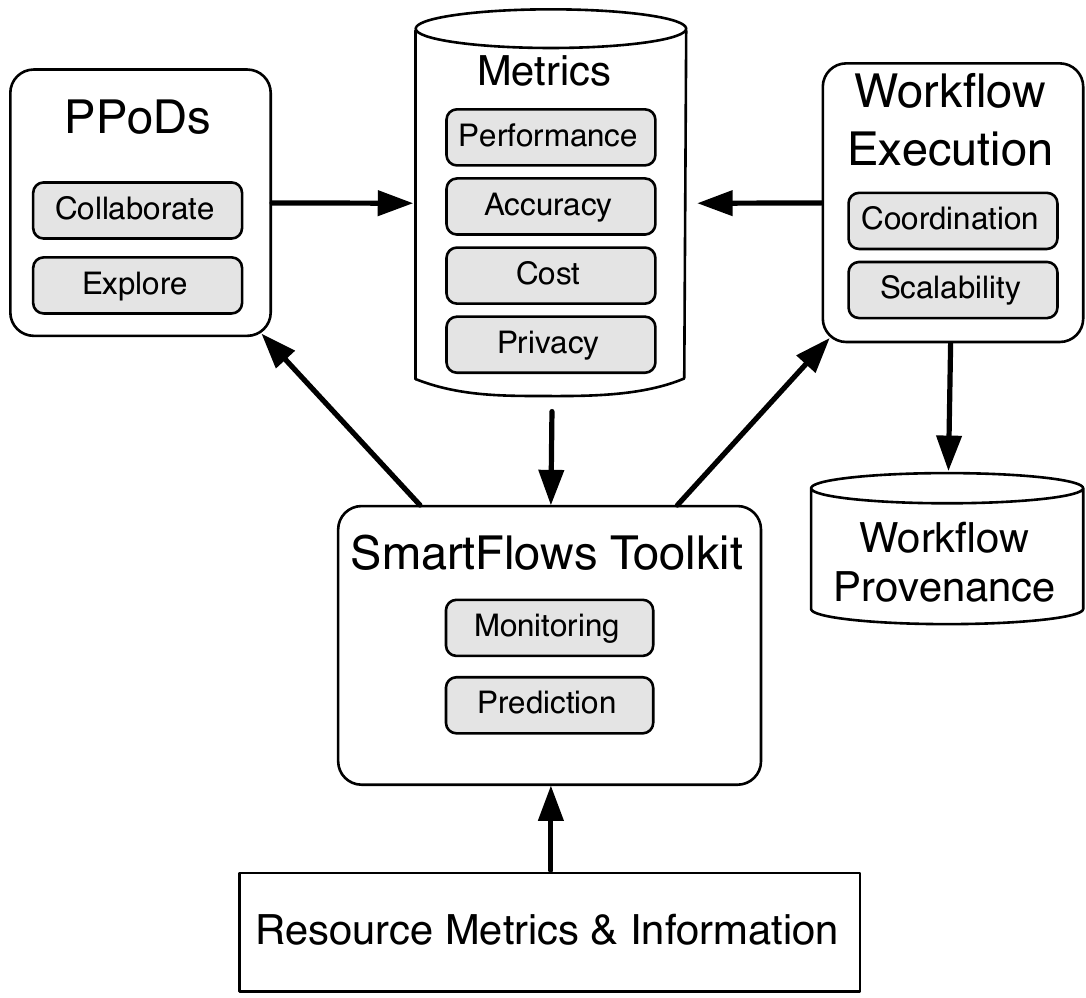}
\caption{The high-level architecture showing the dependencies between the presented collaborative workflow-driven science tools.}
\label{fig:teamsciencearch}
\vspace{-12pt}
\end{figure}   

In this paper, we discuss a cyberinfrastructure ecosystem and tools around workflows for computational team science powered by artificial intelligence. This architecture, illustrated in Figure~\ref{fig:teamsciencearch}, creates an end-to-end support structure to enable teams to \textit{communicate} and \textit{collaborate}, \textit{explore} and \textit{scale}, and \textit{evaluate} scientific work \textit{using a data-driven approach} by providing extensions to workflow systems with dynamic, predictable and programmable interfaces to teams, systems and scalable infrastructure. 

Figure~\ref{fig:teamsciencearch} shows our end-to-end data-driven workflow reference architecture. The main components of the architecture include \textit{PPoDS} for collaboration measurement, task validation and exploration, \textit{SmartFlows} to provide intelligence through analysis of the metrics and any workflow management and scalable execution environment to pass on the insights delivered by SmartFlows, and other data collected through the exploratory activities within PPoDS. These tools are overlayed on any data and computing infrastructure that exploratory and scalable analysis can be executed on, and the performance data, provenance and output from each step can be collected. We will review PPoDS and SmartFlows in detail in the next two sections. 

Note that any workflow system that needs performance-related intelligence on the individual steps of the workflow can be plugged into this framework to take in the exploratory workflow and create a scalable workflow executing on a distributed infrastructure. For this conceptual paper, we will assume both the prototype and actual workflows execute various steps as containers running services, e.g., Jupyter notebooks \cite{jupyter}, through a dynamic resource manager, e.g., Kubernetes \cite{kubernetes}, and collect performance data. 

\section{PPoDS Methodology and Framework}

PPoDS stands for ``Process for the Practice of Data Science''. It is being developed to empower computational data science teams with effective collaboration tools during the exploratory workflow development phase. In this section, we describe the PPoDS methodology and the associated tools under construction. 

\subsection{PPoDS Methodology: Test-Driven Development for Computational Data Science Workflows}

As a methodology, PPoDS grew out of a need to enable cross-disciplinary data science teams to start building a workflow process in meetings. The process built in these meetings is then used to agree upon measurable accountability metrics for each iteration of the exploration and analytical process development activity. 

\begin{figure}
\centering
\includegraphics[width=0.48\textwidth]{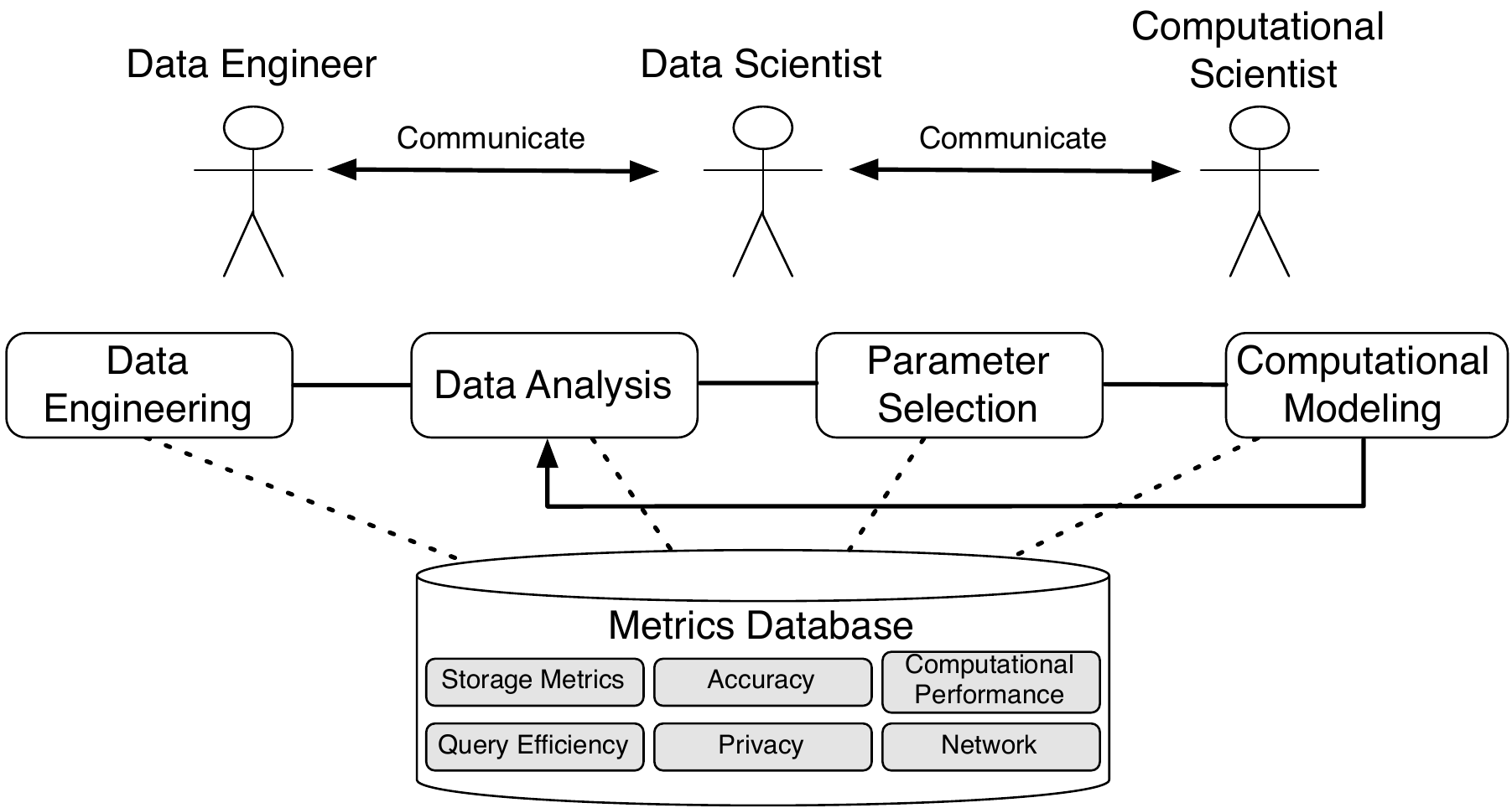}
\caption{Four steps within a conceptual computational data science workflow developed by three collaborating cross-disciplinary team members.}
\label{fig:ppodswf}
\vspace{-12pt}
\end{figure}   

Consider the simple scenario in Figure~\ref{fig:ppodswf}. We have three team members with complementary expertise on data engineering, data analysis and computational science in a domain area. The data engineer makes sure that the experimental data is acquired, modeled and queried effectively for analysis and computational modeling. The data scientist generates insights from the data so that the computational model can be parameterized effectively. The data and computational scientists work together on parameter estimation so that the computational model benefits from the data analysis. Once exploration is complete, the entire workflow needs to run autonomously at scale, on top of a 24x7 experimental data stream. For even further simplification, assume all steps will run within Jupyter notebooks through a scalable environment managed by Kubernetes. This pipeline can be developed as a series of notebooks by different members of the team based on their individual expertise. The development activity starts by agreeing on the steps and the expectation of the team members from each step concerning, e.g., speed, data resolution and quality, accuracy and privacy. The PPoDS methodology suggests that after these metrics are discussed, formalized and recorded as test cases, all team members resume work individually towards these metrics. When everyone reports completing the work that satisfies the developed metrics or raises issues for why it cannot be completed, the team gets together again for another iteration of work. 

To summarize, the cycle of a PPoDS-based workflow development activity is as follows: 
\begin{enumerate}
\item The team develops an understanding of the approach to solve a problem and creates a set of conceptual steps, assigned to team members based on their expertise.
\item The team decides on success metrics and testing approach for each conceptual step.
\item Each member separately develops the steps assigned to them to pass the tests.
\item When each member either reports success (the tests passed) or raised concerns to meet again, everyone creates reports to explain their progress.
\item The team meets again to integrate, share lessons learned and create a consensus report. They also develop new iteration metrics and steps as needed after the the previous iteration. The process goes back to step 3.
\end{enumerate}
This process continues until the team agrees that there is enough exploratory analysis available to start workflow execution at scale. 

\subsection{PPoDS Measurement and Exploration Interface}

After experimenting with the PPoDS methodology, we observed a lack of tools for measuring and testing the development of each individual step in an analytical process towards integration. We are currently developing the tools for capturing, measuring, collecting and analyzing performance metrics during exploratory workflow development and testing process.

Although these tools will eventually be extended to fully support measurable collaboration management, and web-based metric setting and testing for any task within the workflow, we started with a simpler framework.

Our measurement framework is designed to provide web-based container integration and deployment of the developed parametric scientific workflows on customizable infrastructure.
Within this web interface, each container can be treated as an individual composable step and integrated into the overall workflows as a service. Each step is run through the interface and the performance data for the pre-defined metrics gets collected during these exploratory runs. For workflow coordination, we currently use a simple synchronous dataflow pipeline of Jupyter notebooks running through Kepler WebView \cite{webview} on the CHASE-CI through containers. NSF CHASE-CI is a network of fast GPU appliances for machine learning and storage managed through Kubernetes on the high-speed Pacific Research Platform (PRP) \cite{prp}. 

When starting out with a new experiment or workflow it is usually built as a serial process. Steps are tweaked and changed until it is running as expected and then the workflow is scaled out. However, there is sometimes a lot of code refactor that is involved when scaling.  Steps need to be split up in certain ways and inputs and outputs need to be dispersed among worker threads.  It would be better if code could be written such that it would not have to be refactored in order to take advantage of scaling up. This is when the task-based PPoDS methodology and the Explore-to-Scale workflow framework becomes needed.

\subsection{PPoDS Explore-to-Scale Workflow Framework}

To achieve autonomous performance scalability for the integrated workflow without reengineering the exploratory workflow, each step is treated in the workflow as a composable service that gets measured during the exploration through the PPoDS interface. In the case of Jupyter notebooks as individual workflow steps, we treat them as small micro-services that are running some sort of business logic.  The dependencies are stripped away from these micro-services and they work and interact through inputs and outputs of message queues. By utilizing message passing and queueing systems, our approach provides a number of advantages.  The workflows are split up in a way that allows for testing and exploring of different parts of the flow.  Notebooks can be switched in and out of the workflow that would allow for the code to be highly transformable and the ability for changes to be made quickly.

\begin{figure}
\centering
\includegraphics[width=0.48\textwidth]{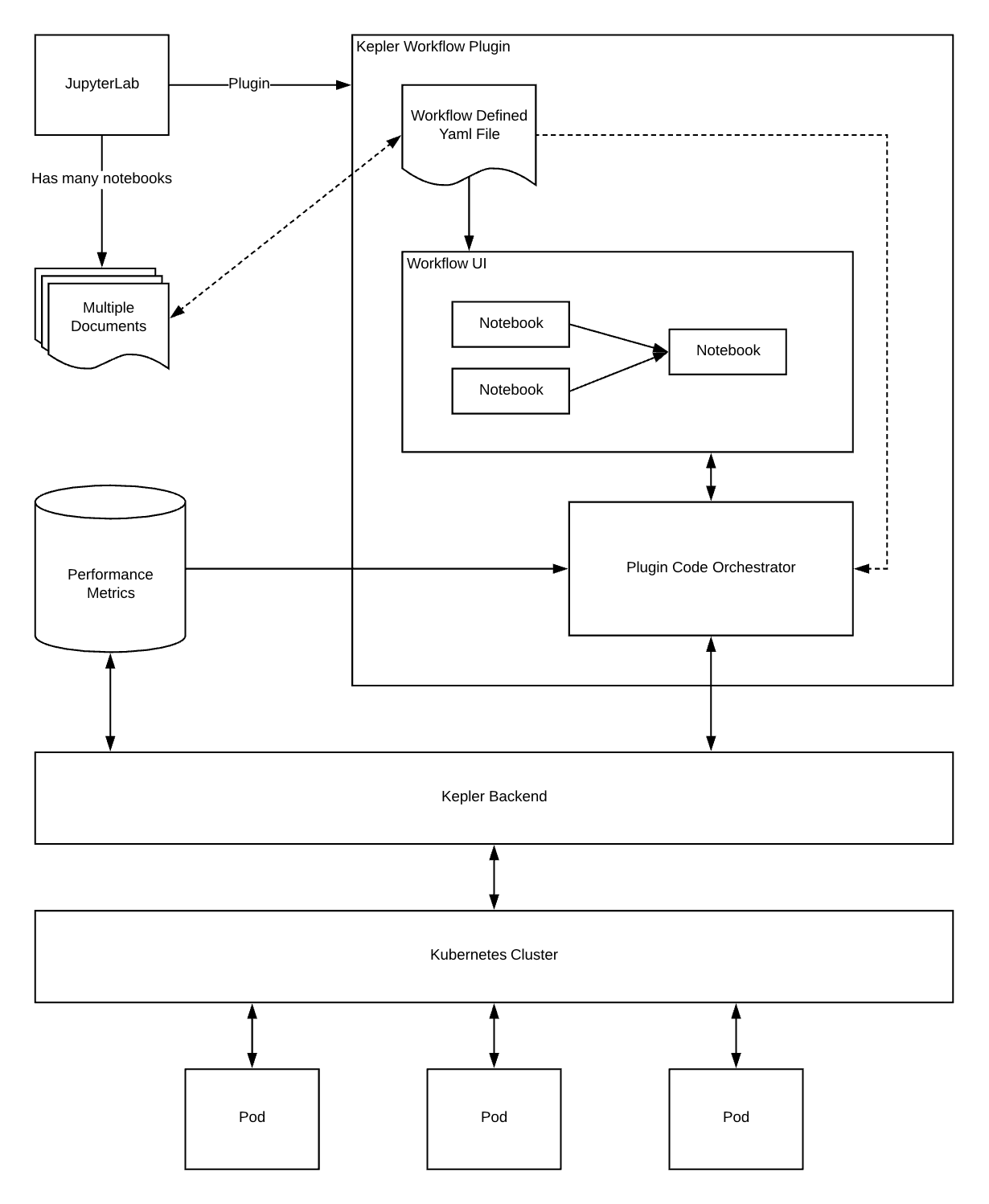}
\caption{Kepler Workflow Plugin for the JupyterLab-based PPoDS Explore-to-Scale web user interface.}
\label{fig:keplerworkflow}
\vspace{-12pt}
\end{figure}   

Figure~\ref{fig:keplerworkflow} shows the architecture for the JupyterLab-based PPoDS explore-to-scale interface using a Kepler workflow plug-in for management of workflows constructed out of Jupyter notebooks. One of our design goals was the system flow to be highly scalable but at the same time very usable and reliable. The barrier to entry is kept as small as possible so that more time is spent on science rather than on debugging code. Note that the Jupyter notebooks should be built in a way that they can communicate with each other through a queuing system. The queuing system acts as a broker for communication between notebooks and provides a routing advantage, giving the system the ability to selectively assemble different notebooks and pass them work.  This can be scaled up to take advantage of many nodes when further scalability is required.

As a simple example, let us consider a machine learning algorithm that the user wants to run. However the whole algorithm is in a single Jupyter notebook that is not built to scale, even if the machine learning portion of the code uses a framework that is scalable given the correct parameters. In order to run at scale, this notebook can be re-written as a series of notebooks that do scalable operations on portions of the code. 
Machine learning begins by gathering the data, cleaning it up and presenting it in a normalized and vectorized format to a framework like TensorFlow \cite{tensorflow}.  A lot of this data pre-processing steps can be migrated to scale to many workers if the input is split up correctly. This is where the message queueing system interacting with worker nodes and the metrics we collected on this notebook during the exploratory phase of the workflow design comes into play. Using the analytical capabilities provided by the SmartFlows Toolkit (see Sec.~\ref{sec:smartflows}) on the metrics collected, the workflow system can gather intelligence on how many worker nodes are needed. These pre-processing notebooks are scaled up and ran in an orchestration framework such as Kubernetes and talk to a middleware layer that hands out work placed in queues.  Notebooks are be able to communicate to other notebooks telling them different inputs and outputs along with telling a notebook when to start after it has finished a critical data pre-processing step.

Doing all of this gives the collaborating team members both explorability and scalability in their experiments.  They are able to work as they normally would in a Jupyter Notebook environment with the tools they are used to. They can explore their workflow and program by quickly using interchangeable pieces.  At the same time, they can begin to explore at scale by using more nodes to run their computations.  Figure~\ref{fig:keplerworkflow} shows the moving pieces in this new workflow design, beginning with JupyterLab and moving all the way down to the pods running in Kubernetes.  we believ this is a scalable architecture that will allow for the design of science to evolve rapidly with code and compute power as the backbone.


\section{SmartFlows Toolkit}
\label{sec:smartflows}

\begin{figure}
\centering
\includegraphics[width=0.35\textwidth]{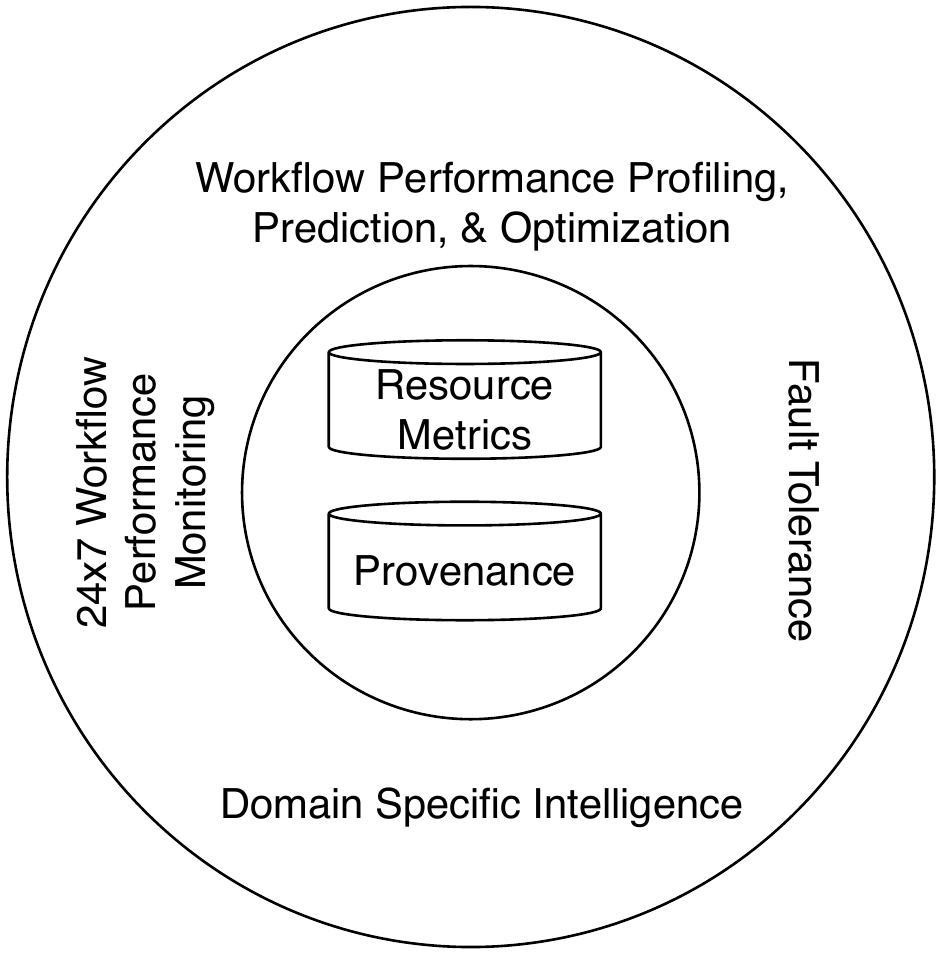}
\caption{Four dimensions of the SmartFlows Toolkit providing analytical services on top of the real-time workflow and infrastructure data: Monitoring, Optimization, Fault Tolerance and Domain Specific Analysis.}
\label{fig:smartflowstoolkit}
\vspace{-12pt}
\end{figure}

The SmartFlows Toolkit is a suite of tools designed to operate on a 24x7 basis for collecting, monitoring and analyzing metrics from the PPoDS exploratory process and underlying infrastructure (Figure \ref{fig:smartflowstoolkit}).  The data driven intelligence provided by SmartFlows analytical services are consumed by the workflow management components to take smart decisions for the current workflow execution or future executions. Here we describe services provided by the SmartFlows toolkit.

\textbf{Workflow Performance Profiling, Prediction and Optimization Services - } The SmartFlows Toolkit  provides services for dynamic analysis of system state and workflow task progress.  Our prior work \cite{Singh1, Singh2} showed leveraging Machine Learning and Deep Learning techniques to predict performance and suggest optimal resources for execution based on workflow applications and input data. The service is now being developed to enable any workflow system to implement Smart Workflow Resource Selection, to predict the performance of a workflow on available resources and provide suggestions to the workflow engine about the best infrastructure to use based on anticipated execution time and resource cost. 

The Smart Workflow Resource Selection service is based on our framework that leverages Machine Learning models for creating precise performance predictions of unseen modules of a workflow \cite{Singh1}. In \cite{Singh2}, we presented the results of the modular resource-centric approach on a compute and data intensive Microbiome Taxonomy and Gene Abundance workflow (MTGA). The SmartFlows services for profiling data collection and performance prediction were utilized for dynamic coordination and resource optimization in this usecase. We are generalizing these performance prediction services using the presented resource and task metrics database to provide insights to enable optimal workflow scheduling on distributed platforms under the constraints provided by users. 

\textbf{24x7 Workflow Performance Monitoring Services  - } We are also developing SmartFlows services to enable users to view and control the progress of workflows from their handheld devices \cite{webview}. This will enhance the possibility of adding human intelligence input to an already running workflow, and give the user the freedom to monitor and steer workflow execution from anywhere. We envision a future where scientists are able to interact with workflow engines from any place. This will empower scientific computing to execute, track and re-run complex experiments anytime, anywhere.
 
\textbf{Fault Tolerance Services - } While it is difficult to anticipate sources of errors in a truly dynamic environment that is shared by many people, a workflow-based approach can empower us to become more resilient through the development and use of data-driven fault tolerance services. SmartFlows provides currently experimental containerized analytical services to improve execution reliability of workflows by taking advantage of the provenance database. The existing experimental services are being extended to constantly monitor the provenance data stream to detect changes in system states that can trigger a fault.

In addition, our current experimental Fault Tolerance Services leverage Machine Learning techniques to forecast failures ahead of time, allowing the workflow engine to preempt failures. To ensure reliability of execution, in \cite{Singh3} and \cite{Singh4}, we used a Deep Learning based system to monitor execution trails and predict any potential roadblocks. We demonstrated that with enough data gathering, it is possible to make real-time predictions of the final states (success or failure) of a dynamic job, thus enabling preventing actions by our workflow engine. This approach does not require information about the internals of the code or the data that is ingested. We applied our Deep Learning-based service on distributed High Energy Physics computing workflows in a dynamic manner and successfully predicted the eventual success or failure of jobs with 85\% accuracy.

The key advantages of predicting failure-prone jobs is the potential for designing intelligent Fault Tolerance mechanisms to handle anomalous events. The decision-making process will be delegated to the framework in the future, in which case, it will perform data analytics on the provenance database to detect failures and dynamically decide best execution route for robust execution of a workflow. This dynamic fault tolerance framework can enable optimal execution strategies in the face of system failures, thus avoiding duplication of efforts, and reducing cost and time of scientific experiments.

\textbf{Domain Specific Intelligent Services -} SmartFlows is designed as an extensible service (i.e., Kubernetes PODs) repository for handling big data analytics and time series analysis for domain specific data coming out of other steps in a workflow. We provide a way to link domain specific analysis as a part of the workflow for tasks concerning data-driven decisions before the next step of the workflow takes place, e.g., parameter and state estimation, data uncertainty quantification and sensitivity analysis. 

As a summary, multiple intelligent SmartFlows services is designed to handle different areas of a workflow, managing the dynamic execution, ensuring fault tolerance, managing domain specific outcomes using parameterization, guaranteeing timely notifications of anomalies, and cost-effective resource utilization. Collectively, the SmartFlows services bring the power of Machine Learning and Deep Learning that intelligently manages the workflow, and informs the user of the workflow progress in a 24x7 manner.

\section{Related Work}
The presented reference architecture for team science is a unique effort that brings together multiple components and tools related to end-to-end workflow development as a whole. To the best of our knowledge, there is no methodology and testing platform for collaborative workflow design similar to the PPoDS methodology and framework we introduce in this paper. Moreover, the presented framework uniquely designed with the "explore-to-scale" mindset, to bridge the exploration phase of workflow development to its scalability phase by collecting workflow metrics and other infrastructure-related information for a variety of workloads. Indeed, building a system to scale is always needed when trying different data experiments and configurations of parameters, but often these phases are handled disjointly by teams. In this section, we review the related work on team science and performance analysis aspects of our work.

\textbf{Team Science - } Recently there has been a lot of work being made to improve the experience of Jupyter Notebooks. The JupyterLab project has redefined the notebook experience and improved the user interface. The new interactive workspace allows for customizations to the active notebook area that could potentially allow for simpler collaboration between notebooks. JupyterHub \cite{jupyter} and Binder \cite{binder} have also shown how team science can work together in larger groups by allowing them to deploy notebooks on demand and at scale.

Netflix is another innovator in this space that is taking the Jupyter Notebooks to the next level \cite{netflix}.  They rely heavily on big data and machine learning to make fast, quick decisions about what customers might want to watch next.  In order to build these models, data science is a main priority involving many collaborative engineering projects.  Since Netflix is so concerned with scale and experimentation of data, they have built brand new ways that data engineering teams can interact with each other and their code.  They used the concept of a notebook template for code reusability, this would allow a notebook to take in different parameters and dynamically change.  Developing multiple scaling architecture components for notebooks, they were able to scale up their experiments and deploy notebooks as jobs to large computer clusters.

Google is also another big player in this space with all the data that they process. They want to give data scientists the ability to collaborate and bring their experiments to life faster through computation. Using their Colaboratory service, they allow notebooks to run on Google servers and use a GPU or TPU \cite{colaboratory}. It also adds sharing of notebooks through Google Drive and commenting on different cells. This ability to share code and data increases interaction for every one involved in designing a notebook.

There has also been efforts to define team science effectiveness from a social, organizational process and theoretical perspectives \cite{NASSOS, StrengthNumbers}. Although these are very important studies for understanding collaborative team science, it is outside the scope of this paper, in which we explore the measurable operational aspects of team science and explore how workflow-driven thinking can help with it. 

\textbf{Workflow Performance Monitoring, Analysis and Prediction - } Distributed workflows have matured as powerful tools for scientists to develop and deploy large scale scientific experiments on platforms that were previously only available to experts in computing. 
As this trend grows, the room for inefficient scheduling reduces, especially so on multi-tenant platforms with resource contentions. 

Many developments in performance prediction can be categorized into Regression \cite{Marin, Chatzopoulos, Singh1, Singh2,Ferreira}, Classification \cite{Delimitrou, Pumma}, Similarity based predictions \cite{Li, Smith}, and dynamic Time-Series predictions\cite{Zhang, Yang1, Yang2, Singh3, Singh4, Wolski1} . We can also index these techniques on other dimensions, depending on priority of workflow application, such as ease of data collection for Machine Learning techniques, level of sophistication involved in instrumenting a resource for time-series based dynamic prediction, source-code evaluation for intrusive methods, and workload modeling methods for platforms with resource-contention.
 
Regression techniques \cite{Marin},\cite{Chatzopoulos}, \cite{Singh1}, \cite{Singh2}, \cite{Ferreira} model the prediction in terms of discovery of a function or deploy complex machine learning techniques that require large amounts of training data.  ML based methods have shown reliable results for complex workflows and demonstrate potential to scale, but require a robust data-collection and instrumentation pipeline for training data collection. \cite{Singh1}, \cite{Singh2} present a scalable prediction framework that uses divide-and-conquer method to break a large workflow in modules. The paper uses Machine Learning agents for prediction, but the framework can leverage other low-level techniques, due to its modular design. \cite{Marin} uses low-level information and characterizes code binaries to model application-architecture relationships, predicting cache behavior and execution times. \cite{Chatzopoulos} investigates patterns in CPU wait times, and extrapolates them to large machines with many cores. However, \cite{Chatzopoulos} focuses only on in-memory applications.  \cite{Ferreira} presents an on-line estimation scheme that uses correlations among data sets  combined with clustering techniques for non-obvious scenarios, where correlations are hard to model. \cite{Pumma} deploys application-profiles to estimate run times, and tests on real data sets. This work samples workload, performs classification and leverages category specific model for run time prediction. 

Researchers have exploited instance-based learning methods \cite{Li}, \cite{Smith} for estimating durations of file transfer and execution.  The estimation is based on a similarity measure, such as a distance function that extracts `similar' footprints from existing datasets. \cite{Zhang} estimates execution time on grids by mapping the problem to CPU load estimation task, and deploys similar historic time series patterns to correct for polynomial estimation errors. \cite{Singh3}, \cite{Singh4} demonstrate efficacy of neural-network and Machine Learning based methods to provide precise time-series predictions, when such dynamic data is available. Such techniques require instrumentation infrastructure that produce dynamic sensor-generated signals capturing the current state of cyber-infrastructure.

Robust large-scale workflow execution engines of the future will have capacity to generate precise predictions and dynamically evaluate critical performance estimates that feed resource allocation layers. Workflow execution engines with in-built prediction capability can significantly accelerate workflow tasks higher up in the dependency chains, unlocking large-scale cost-savings.


\section{Conclusion}
In this paper, we present an end-to-end architecture that is streamlined to scale exploratory science in a seamless and efficient way. We argue that scientific workflows will provide critical components in accelerating the future of science, and take a central role in the collaborative process of innovation. The paper presents PPoDS methodology that is designed to reduce problems arising due to lack of communication, by enforcing a test-driven development cycle in science. The SmartFlows provides an automated toolkit for data-driven intelligence. Using our framework, scientists will be able to exploit advances in Machine Learning and Internet of Things to obtain actionable insights that enable smarter decisions, based on dynamically measured data. 
By re-thinking the design and discovery process, we present a methodology that aims to first and foremost facilitate effective collaboration, provide inbuilt fault-tolerance and reliability to workflows of future, drastically reduce execution bottlenecks by constantly measuring, learning, and informing every aspect of a scientific workflow.

We did not present experimental results as it was necessary to put together this architecture conceptually before any of the individual components in the architecture could be presented. Why and how different pieces come together required the conceptual nature of the approach presented in this paper. We hope that the presented figures and architecture diagram gives the reader what is being developed and opens up opportunities for discussions leading to better collaboration tools that lead to intelligent scalable workflows.

Beyond all, this paper presented a motivation for the importance of thinking computational data science as a whole ecosystem including people, processes and systems, all of which can be measured and optimized using a workflow-driven approach. We believe there is an opportunity to carry workflows from being performance optimization and task orchestration tools to acting as intelligent operational research tools in computational data science conducted by teams.


%

\section*{Acknowledgments}

This work was supported in part by NSF-1331615 under CI, Information Technology Research, and SEES Hazards programs for WIFIRE, DOE DE-SC0012630 for IPPD, NSF-1730158 for Cognitive Hardware and Software Ecosystem Community Infrastructure (CHASE-CI), and NIH P41 GM103426 for National Biomedical Computation Resource (NBCR). The content is solely the responsibility of the authors and does not necessarily represent the official views of the funding agencies. The authors would also like to thank Workflows for Data Science (WorDS) Center of Excellence team members Mai Nguyen and Volkan Vural for their participation in the discussions leading to this paper.


\begin{IEEEbiography}{Ilkay Altintas}
Dr. Ilkay Altintas is the Chief Data Science Officer and the Director for the Workflows for Data Science (WorDS) Center of Excellence at the San Diego Supercomputer Center, UCSD. Since joining SDSC in 2001, she has worked on different aspects of scientific workflows as a principal investigator and in other leadership roles across a wide range of cross-disciplinary projects. She is a co-initiator of and an active contributor to the open-source Kepler Scientific Workflow System, and the co-author of publications related to computational data science at the intersection of scientific workflows, provenance, distributed computing and software modeling. Ilkay received a Ph.D. degree from the University of Amsterdam, the Netherlands, with a focus on provenance of workflow-driven collaborations.
\end{IEEEbiography}

\begin{IEEEbiographynophoto}{Shweta Purawat}
Shweta Purawat is New-User Applications Specialist for the WorDS Center of Excellence at the San Diego Supercomputer Center, UCSD. She develops compute and data intensive workflows and actors for distributed platforms that involve cloud and grid environments. She is excited about Cloud Computing, Data Science Workflows, Scientific Workflow design, Large Scale Data Intensive Computing, and Predictive Analytics. Her interest lies in designing core technology tools and applications, that act as catalyst for innovation. She holds M. Tech degree from IIT Bombay, India. Prior to joining SDSC, Shweta was a Design Engineer at Intel Corporation.
\end{IEEEbiographynophoto}


\begin{IEEEbiographynophoto}{Daniel Crawl}
Dr. Daniel Crawl is the Associate Director for the WorDS Center of Excellence at the San Diego Supercomputer Center, UCSD. He is the lead architect for the overall integration of distributed data parallel (DDP) execution patterns and the Kepler Scientific Workflow System. He conducts research and development of execution patterns, bioActors, and distributed directors. Daniel received a Ph.D. in Computer Science from the University of Colorado.
\end{IEEEbiographynophoto}

\begin{IEEEbiographynophoto}{Alok Singh}
Alok Singh research interests are Predictive Modeling, Machine Learning, Neural Networks, Data Management Technologies, Large Scale Data Analytics, and Distributed Data Processing. Alok graduated with a MS in Computer Science at UCSD, and has B.Tech and M.Tech degrees from IIT Bombay, India. 
\end{IEEEbiographynophoto}

\begin{IEEEbiographynophoto}{Kyle Marcus}
Kyle is a software engineer and researcher in the WorDS Center of Excellence at the San Diego Supercomputer Center, UCSD. His research interests include HPC, cloud computing and programming languages. In previous industry experience, Kyle has worked on large scale distributed systems and embedded devices. When not programming, he can be found indulged in many of his hobbies including astrophotography and 3d printing. He graduated with a MS in Computer Science from the University at Buffalo.
\end{IEEEbiographynophoto}

\end{document}